\documentclass[10pt]{article}
\def\Fig#1{Fig.~\ref{fig:#1}}

\def\Figspace#1#2#3{}%\begin{figure}
\def\Eq#1{Eq.~(\ref{eq:#1})}

\def\Eqnum#1{(\ref{eq:#1})}
\def\beginEq{\begin{equation}}
\def\endEq{\end{equation}}
\def\beginEqarray{\begin{eqnarray}}
\def\endEqarray{\end{eqnarray}}
\def\labelEq#1{\label{eq:#1}}

\def\bra#1{\langle {#1} | }
\def\ket#1{| {#1} \rangle }
\def\braket#1#2{\langle {#1} | {#2} \rangle }
\def\expval#1#2#3{\bra {#1} {#2} \ket {#3} }
\def\avg#1{\langle {#1} \rangle }

\def\nl{\nonumber \\}

\def\O{{\cal O}}		    % script O for order of
\def\Lag{{\cal L}}    % script L for Lagrangian
\def\P{{\cal P}}      
\def\Q{{\cal Q}}     
\def\eff{{\rm eff}}   % effective
\def\H{{\cal H}}

\def\G2b{\overline{G^2}}
\def\psib{\overline{\psi}}
\def\G{{\cal G}}

\def\Eb{\overline{E}}
\def\ext{{\rm ext}}
\def\GeV#1{$#1\:$GeV}

\def\pn1p{{(n+1)}}
\def\equal{\! = \! }
\def\greater{\! > \! }

\def\ub{\overline{u}}
\def\Dv{{\vec D\,}}
\def\sigmav{{\vec\sigma\,}}
\def\Ev{{\vec E\,}}
\def\Bv{{\vec B\,}}
\def\psid{{\psi^\dagger}}
\def\delv{{\nabla}}

\def\half{\mbox{$\frac{1}{2}$}}
\newenvironment{Exercise}{\small\begin{description}
                         \item[Exercise:]}{\end{description}}
\newenvironment{Ref}{\small\begin{quote}}{\end{quote}}

\usepackage{feynmp}
\def\enddoc{\end{document}}

\title{What is Renormalization?}
\author{G.Peter Lepage \\ \small Newman Laboratory of Nuclear Studies \\ 
        \small Cornell University, Ithaca, NY 14853}
\date{\small Talk presented at TASI'89, June 1989}

\begin{document}
 
\maketitle
 
\begin{fmffile}{fdiags}

 \section{Introduction}

As everyone knows the quantized theory of electrodynamics was created in the
late 1920's and early 1930's. The theory was analyzed in perturbation theory,
and was quite successful to leading order in the fine-structure constant
$\alpha$. However all sorts of infinities started to appear in
calculations beyond that order, and it was almost twenty years before the
technique of renormalization was developed to deal with these infinities. What
resulted is quantum electrodynamics (QED), one of the most accurate physical
theories ever created: the $g$-factor of the electron is predicted (correctly)
by the theory to at least 12 significant digits!  At first sight, 
renormalization appears to be a rather dubious procedure for
hiding embarrassing infinities, and the success of QED seems nothing less than
miraculous. Nevertheless, persuaded by success, most physicists decided that
renormalizability was an essential ingredient in any physically relevant field
theory. Such thinking played a crucial role first in the development of a
fundamental theory of weak interactions and then in the discovery of the
underlying theory of strong interactions. In this lecture I argue
that renormalizability is {\em not} an essential characteristic of useful field
theories. Indeed it is possible, some would say likely, that none of known
interactions is described completely by a renormalizable
field theory. Modern developments in renormalization theory have given meaning
to nonrenormalizable field theories, thereby generalizing and greatly clarifying
our understanding of quantum field theories. As a result nonrenormalizable
interactions are possible and seem likely in most theories constructed to deal
with the real world.

In the first part of this lecture we will examine the technique of
renormalization, first illustrating the conventional ideas and then extending
these to deal with nonrenormalizable interactions. Central to this discussion is
the notion of a cut-off field theory as a low-energy approximation to some more
general (and possibly unknown) theory. Much of this material warrants a more
detailed discussion than we have time for, and so a number of exercises have
been included to suggest topics for further thought.

In the second part of the lecture we will examine the implications of our new
perspective on renormalization for theories of electromagnetic, strong, and weak
interactions. Here we will address such issues as the origins and significance of
renormalizability, the importance of naturalness in physical theories,
and the experimental limits on nonrenormalizable interactions in electromagnetic
and weak interactions. We will also show how to use renormalization ideas to
create rigorous nonrelativistic field theories that greatly simplify the analysis
of such nonrelativistic systems as positronium or the $\Upsilon$ meson.

Most of the ideas presented in this lecture are well known to many people.
However, since little of the modern attitude towards renormalization has made it
into standard texts yet, it seems appropriate to devote a lecture to the
subject at this Summer School. The discussion presented here is largely
self-contained; the annotated Bibliography at the end lists a few references that
lead into the large literature on this diverse subject.

\section{Renormalization Theory}

\subsection{The Problem with Quantum Fields}

The infinities in quantum electrodynamics, for example, originate in the fact
that the electric field $E(x,t)$ becomes a quantum-mechanical operator in the
quantum theory. Thus measurements of $E(x,t)$ in identically prepared systems
tend to differ: the electric field has quantum fluctuations. By causality,
adjacent measurements of the field, say at points $x$ and $x+a$, are independent
and thus fluctuate relative to one another. As a result the quantum
electric field is rough at all length scales, becoming infinitely rough at
vanishingly small length scales.

 \begin{Exercise}
Define $\Eb(x,t)$ to be the electric field averaged over a spherical region of
radius $a$ centered on point $x$. This might be roughly the field measured by a
probe of size $a$. Show that 
 \beginEq
\expval{0}{\left(\Eb(x+a,t)-\Eb(x,t)\right)^2}{0} \to \frac{1}{a^4}
 \endEq
as $a\!\to\! 0$---i.e., the fluctuations in the field from point to point 
diverge as the probe size goes to zero, even for the vacuum state!
Physically, the fluctuations arise because it is impossible to probe the
electric field at a point without creating photons.
 \end{Exercise}

This structure at all length scales is characteristic of quantum fields, and is
quite different from the behavior of classical fields which typically become
smooth at some scale. The roughness of the field is at the root of the problem
with defining the quantum field theory. For example, how does one define
derivatives of a field $E(x)$ when the difference $E(x+a)-E(x)$ diverges as the
separation $a$ vanishes? In perturbation theory the roughness at short distances
results in divergent integrations over loop momenta, divergences associated with
intermediate states carrying arbitrarily large momenta (and having arbitrarily
short wavelengths). The infinities that result demonstrate rather
dramatically that the short-distance structure of the quantum fields plays an
important role in determining the long-distance (low-momentum) behavior of the
theory; the short-distance structure cannot be ignored. 

To give any meaning at all to a quantum field theory one must first regulate it,
by in effect removing from the theory all states having energies much larger than
some cutoff $\Lambda$. With a cutoff in place one is no longer plagued by
infinities in calculations of the scattering amplitudes and other properties of
the theory. For example, integrals over loop momenta in perturbation theory are
cut off around $\Lambda$ and thus are well defined. However the cutoff seems very
artificial. The use of a cutoff apparently contradicts the notion, developed
above, that the short-distance structure of the theory is important to the
long-distance behavior; with the cutoff one is throwing away the short-distance
structure. Furthermore $\Lambda$ is a new and artificial parameter in the
theory. Thus it is traditional to remove the cutoff by taking $\Lambda$ to
infinity at the end of any calculation. This last step is the source of much of
the mystery in the renormalization procedure, and it now appears likely that this
last step is also a wrong step in the nonperturbative analysis of many theories,
including QED. Rather than follow this route we now will examine what it means to
keep the cutoff finite.

\subsection{Cut-off Field Theories}

The basic idea behind renormalization is that all effects of the very high-energy
states in the Hilbert space on the low-energy behavior of the theory can be
simulated by a set of new local interactions. So we can discard the states with
energy greater than some cutoff provided we modify the theory's Lagrangian to
account for the effects that result from the discarded states. In this section
we will see how this idea provides the basis for the conventional renormalization
procedure, using QED as an example of a renormalizable field theory. This will lay
the groundwork for our discussion, in the next section, of the origins and
significance of nonrenormalizable interactions.

According to conventional renormalization theory QED is defined by a
Lagrangian,
 \beginEq
\Lag_0 = \psib(i\partial\cdot\gamma - e_0 A\cdot\gamma - m_0) \psi -
\half (F^{\mu\nu})^2,
 \endEq
together with a regulator that truncates the theory's state space at some
very large $\Lambda_0$.\footnote{
 The simplest way to regulate perturbation theory is to simply cut off the
integrals over loop momenta at $\Lambda_0$. Although such a regulator can
(and has been) used, it complicates practical calculations because it is
inconsistent with Lorentz invariance and gauge invariance. However the details
of the regulator are largely irrelevant to our discussion, and so for
simplicity we will speak of the regulator as though it is a simple cutoff. One
of the more conventional regulators, such as Pauli-Villars or lattice
regulators, is recommended for real calculations.
 }
The cut-off theory is correct up to errors of $\O(1/\Lambda_0^2)$. It is worth
emphasizing that $e_0$ and $m_0$ are
well-defined numbers so long as $\Lambda_0$ is kept finite; in QED each can be
specified to several digits (for any particular value of $\Lambda_0$). Given
these ``bare'' parameters one need know nothing else about renormalization in
order to do calculations. One simply computes scattering amplitudes, cutting all
loop momenta off at $\Lambda_0$ and using the bare parameters in propagators and
vertices. The renormalization takes care of itself automatically. To compute
$e_0$ and $m_0$ for a particular $\Lambda_0$ one chooses two convenient processes
or quantities, computes them in terms of the bare parameters using $\Lag_0$, and
adjusts the bare parameters until theory and experiment agree. Then all other
predictions of the theory will be correct, up to errors of
$\O(1/\Lambda_0^2)$.\footnote{
 Newcomers to field theory sometimes find it hard to
believe that this is all there is to conventional renormalization, given the
length and complexity of the treatment generally accorded the subject in texts.
What happened to counterterms, subtractions points, and normalization
conditions? These are all related to the detailed implementation of the general
concepts we are discussing. Such implementations tend to be highly optimized for
particular sorts of calculations---e.g., for high-order calculations in
perturbation theory, or for lattice simulations---and as such can be fairly
complex. Such details are important in actual calculations. Here however our
focus is on conceptual issues and so we can dispense with much of the detail.
Anyone planning to do real calculations is advised to consult the standard texts.
 }

To understand the role of the cutoff in defining the theory, we start with QED
as defined by $\Lag_0$ and cutoff $\Lambda_0$, and we remove from this theory
all states having energies or momenta larger than some new cutoff $\Lambda$
($\ll\!\Lambda_0$). Then we examine
how $\Lag_0$ must be changed to compensate for this further truncation of the
state space. Of course the new theory that results can only be useful for
processes at energies much less than $\Lambda$, and so we restrict our
attention to such processes. Furthermore we will analyze the effects of the new
cutoff using perturbation theory, although our results are valid
nonperturbatively as well.

We now want to discard all contributions to the theory coming from loop momenta
greater than the new cutoff $\Lambda$. Consider first, the one-loop
radiative corrections to the amplitude for an electron to scatter off an external
electromagnetic field. 
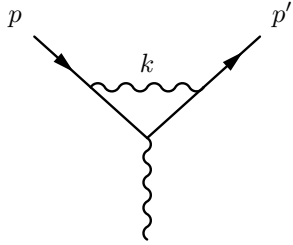
\begin{figure}
\begin{center}
\begin{fmfgraph*}(30,30)
	\fmfset{arrow_len}{3mm}
	\fmftop{i1,o1}
	\fmfbottom{s1}
	\fmf{fermion}{i1,v1} \fmflabel{$p$}{i1}
	\fmf{plain}{v1,v2}
	\fmf{plain}{v2,v3}
	\fmf{fermion}{v3,o1} \fmflabel{$p^\prime$}{o1}
	\fmf{photon}{s1,v2}
	\fmf{photon,tension=0,label=$k$,label.side=left}{v1,v3}
\end{fmfgraph*}
\end{center}
\caption{\label{fig:ver}The one-loop vertex correction to the
          amplitude for an electron scattering off an external field.}
\end{figure}

% \Figspace{ver}{1.36in by 1.04in}{The one-loop vertex correction to the
%          amplitude for an electron scattering off an external field.}
Working in the $\Lag_0$ theory with the original cutoff, the
part of the vertex correction (\Fig{ver}) that is being discarded is
 \beginEqarray
\lefteqn{T^{(a)}(k>\Lambda) = -e_0^3
\int_\Lambda^{\Lambda_0} \frac{d^4k}{(2\pi)^4} \frac{1}{k^2}  \times }\nl
 & & \times \ub(p^\prime) \gamma^\mu \frac{1}{(p^\prime-k)\cdot\gamma-m_0}
A_\ext(p^\prime-p)\cdot\gamma \frac{1}{(p-k)\cdot\gamma-m_0} \gamma_\mu u(p).
 \labelEq{ver}
 \endEqarray
Since the masses and external momenta are assumed to be much less than
$\Lambda$, we can neglect $m_0$, $p$ and $p^\prime$ in
the integrand as a first approximation, thereby greatly simplifying the integral:
 \beginEqarray
T^{(a)}(k>\Lambda) &\approx & -e_0^3 
\int_\Lambda^{\Lambda_0} \frac{d^4k}{(2\pi)^4} \frac{1}{k^2}\;
\ub(p^\prime) \gamma^\mu \frac{k\cdot\gamma}{k^2}
A_\ext(p^\prime-p)\cdot\gamma \frac{k\cdot\gamma}{k^2} \gamma_\mu u(p) \nl
& \approx &
-e_0^3 \,\ub(p^\prime)\,A_\ext(p^\prime-p)\cdot\gamma\,u(p)\;
\int_\Lambda^{\Lambda_0} \frac{d^4k}{(2\pi)^4} \frac{1}{(k^2)^2} .
 \endEqarray
Applying a similar analysis to the other one-loop corrections, we find that the
part of the electron's scattering amplitude that is omitted as a result of the
new cutoff has the form
 \beginEq
T(k>\Lambda) \approx -i e_0\,c_0(\Lambda/\Lambda_0)\:\ub(p^\prime)\, 
A_\ext(p^\prime-p)\cdot\gamma\,u(p)
 \endEq
where $c_0$ is dimensionless and thus can depend only upon the ratio
$\Lambda/\Lambda_0$, these being the only scales left in the loop integration. 

 \begin{Exercise}
Show that 
 \beginEq
c_0(\Lambda/\Lambda_0) = -\frac{\alpha_0}{6\pi} \log(\Lambda/\Lambda_0) .
 \endEq
 \end{Exercise}

Clearly $T(k\greater\Lambda)$ is an important contribution to the electron's
scattering amplitude; it cannot be dropped. However such a contribution can be
reincorporated into the theory by adding the following new interaction to
$\Lag_0$:
 \beginEq
\delta\Lag_0 = -e_0\,c_0(\Lambda/\Lambda_0)\;\psib\, A\cdot\gamma\,\psi .
 \labelEq{de}
 \endEq
Thus we can modify the Lagrangian to compensate for the removal of the states
above the cutoff, at least for the purpose of computing the electron's scattering
amplitude. 

It is important that the new
interaction $\delta\Lag_0$ is completely specified by a single number, the
coupling constant $c_0$. In analyzing the vertex
correction to electron scattering (\Eq{ver}), we can neglect the external
momenta relative to the internal momentum $k$ with the result that
the coupling $c_0$ is independent of the external momenta. Thus the interaction
is characterized by a number, rather than by some complicated function of the
external momenta. Momentum independence, or more generally polynomial dependence
on external momenta, indicates that these effective interactions are local in
coordinate space; that is, they are polynomial in fields or derivatives of the
fields all evaluated at the same point $x$. This important result actually follows
from the uncertainty principle and is quite general: interactions involving
intermediate states with momenta greater than the cutoff are local as far as the
(low-energy) external particles are concerned.  Since by assumption the  external
particles have momenta far smaller than $\Lambda$, intermediate states above the
cutoff must be highly virtual. In quantum mechanics a state can be highly virtual
provided it is short-lived, and so these highly-virtual intermediate states can
exist for times and propagate over distances of only $\O(1/\Lambda)$. Such
distances are tiny compared with the wavelengths of the external
particles, $\lambda \sim 1/p \gg 1/\Lambda$, and thus the interactions are
effectively local. 

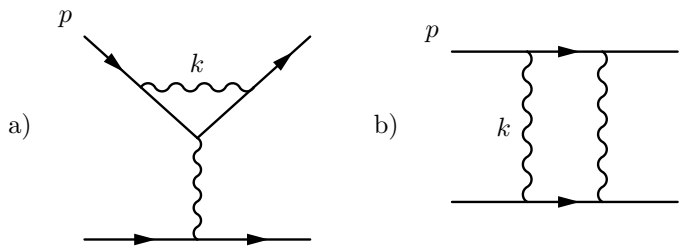
\begin{figure}
\begin{center}
a)\qquad \parbox[c]{0.25\linewidth}{
\begin{fmfgraph*}(30,30)
	\fmftop{i1,o1}
	\fmfstraight
	\fmfset{arrow_len}{3mm}
	\fmfbottom{i2,s1,o2}
	\fmf{fermion}{i1,v1} \fmflabel{$p$}{i1}
	\fmf{plain}{v1,v2}
	\fmf{plain}{v2,v3}
	\fmf{fermion}{v3,o1} %\fmflabel{$p^\prime$}{o1}
	\fmf{photon}{s1,v2}
	\fmf{photon,tension=0,label=$k$,label.side=left}{v1,v3}
	\fmf{fermion}{i2,s1}
	\fmf{fermion}{s1,o2}
\end{fmfgraph*}}
\qquad b)\qquad \parbox[c]{0.25\linewidth}{
\begin{fmfgraph*}(30,20)
	\fmfstraight
	\fmfset{arrow_len}{3mm}
	\fmftop{t1,t2,t3,t4} \fmflabel{$p$}{t1}
	\fmfbottom{b1,b2,b3,b4}
	\fmf{plain}{t1,t2}
	\fmf{fermion}{t2,t3}
	\fmf{plain}{t3,t4}
	\fmf{plain}{b1,b2}
	\fmf{fermion}{b2,b3}
	\fmf{plain}{b3,b4}
	\fmf{boson,label=$k$,label.side=right}{t2,b2}
	\fmf{boson}{t3,b3}
\end{fmfgraph*}}
\end{center}
\caption{\label{fig:ee}Examples of one-loop 
radiative corrections to electron-electron scattering.}
\end{figure}

Although electron scattering on external fields has been fixed, we must
worry about the effect of the new cutoff on other processes. Consider for example
the one-loop corrections to electron-electron scattering (\Fig{ee}).
% \Figspace{ee}{4.07in by 0.92in}{Examples of one-loop radiative corrections to
%          electron-electron scattering.}
The $k\greater\Lambda$ contributions due to vertex and self-energy
corrections to the one-photon exchange process (e.g., \Fig{ee}a) are correctly
simulated by $\delta\Lag_0$, just as they are in the case of an electron
scattering on an external field. That leaves only the $k\greater\Lambda$
contribution from the two-photon exchange diagrams (e.g., \Fig{ee}b). Again one
can neglect the external momenta and masses in the internal propagators. The
resulting amplitude must involve spinors for each of the external electrons,
in combinations like $\ub\gamma_\mu u\,\ub\gamma^\mu u$ or $\ub u\,\ub u$.
These all have the dimension of $[\mbox{energy}]^2$, while in general a
four-particle amplitude must be dimensionless. Thus the part of the amplitude
involving loop momenta larger than $\Lambda$ contributes something like
 \beginEq
 d(\Lambda/\Lambda_0) \,\frac{\ub\gamma_\mu u\,\ub\gamma^\mu
u}{\Lambda^2} ,
 \labelEq{eeee}
 \endEq
where $d$ is dimensionless and where the factor in the denominator is
$\Lambda^2$, since $\Lambda$ is the only important scale left in the loop
integral. Clearly such a contribution is suppressed by $(p/\Lambda)^2$ and can be
ignored (for the moment) since we are assuming $p\!\ll\!\Lambda$.  

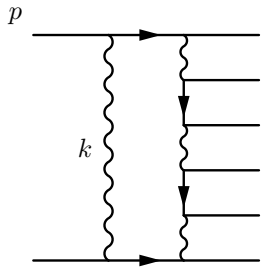
\begin{figure}
\begin{center}
\begin{fmfgraph*}(30,30)
\fmfset{arrow_len}{3mm}
\fmfstraight
\fmftop{t1,t2,t3,t4} \fmflabel{$p$}{t1}
\fmfbottom{b1,b2,b3,b4}
\fmfleft{b1,t1}
\fmfright{b4,r4,r3,r2,r1,t4}
\fmf{plain}{t1,t2}
\fmf{plain}{t3,t4}
\fmf{plain}{b1,b2}
\fmf{plain}{b3,b4}
\fmf{fermion}{t2,t3}
\fmf{fermion}{b2,b3}
\fmf{fermion}{i1,i2}
\fmf{fermion}{i3,i4}
\fmf{boson,label=$k$,label.side=right}{t2,b2}
\fmf{boson}{t3,i1}
\fmf{boson}{i2,i3}
\fmf{boson}{i4,b3}
\fmffreeze
\fmf{plain}{i1,r1}
\fmf{plain}{i2,r2}
\fmf{plain}{i3,r3}
\fmf{plain}{i4,r4}
\end{fmfgraph*}
\end{center}
\caption{\label{fig:ee2}A one-loop radiative correction to electron-electron scattering i
into four electrons and two positrons.}
\end{figure}

A similar analysis for,
say, electron-electron scattering  into four electrons and two positrons
(e.g., \Fig{ee2}) shows that intermediate states above the cutoff contribute
something of order
 \beginEq
\frac{(\ub\gamma u)^2\,(\ub\gamma v)^2}{\Lambda^8} ,
 \endEq
which is even less important. Evidently the more external particles involved in a
loop, the larger the number of hard internal propagators, and the smaller
the effect of the cutoff.
% \Figspace{ee2}{1.08in by 1.14in}{A one-loop radiative correction to
%           electron-electron scattering into four electrons and two positrons.}

 \begin{Exercise}
Show that an amplitude with $n$ external particles has dimension
$[\mbox{energy}]^{4-n}$ when relativistic normalization (i.e.,
$\braket{p\ldots}{p^\prime\ldots}\propto \sqrt{p^2+m^2}\,\delta^3(p-p^\prime)$)
is used for all states. Use this fact to show that the addition of an
extra pair of external fermions to a $k\greater\Lambda$ loop results in
an extra factor of $1/\Lambda^{3}$, while adding an extra external photon
leads to an extra factor of $1/\Lambda$. For some processes additional factors of
external momenta or of the electron mass may also be required respectively by
gauge invariance or chiral symmetry (i.e., electron-helicity conservation when
$m\equal0$). Such factors result in additional factors of
$1/\Lambda$.
(Note that the relativistic normalization condition for
Dirac spinors is $\ub u\equal 2m$, and that photon polarization vectors are
normalized by $\varepsilon\cdot\varepsilon = 1$.)
 \end{Exercise} 

 Using simple dimensional and power-counting arguments of this sort,
one can show that the only scattering amplitudes that are strongly affected by 
the cutoff are those involving the electron-photon vertex, and for these the
loss of the $k\greater\Lambda$ states is correctly compensated by adding the
single correction $\delta\Lag_0$ to the Lagrangian. The only other physical
quantity that is strongly affected by the cutoff is the mass of the electron: the
physical mass of the electron is $m_0$ plus a self-energy
correction that involves the $k\greater\Lambda$ states. The effect of these
states on the mass is easily simulated by adding a term of the form
 \beginEq
- m_0\,\tilde{c}_0(\Lambda/\Lambda_0)\,\psib \psi
 \labelEq{dm}
 \endEq
to the Lagrangian.\footnote{
 It is not obvious at first glance that this mass correction is proportional to
$m_0$. On strictly dimensional grounds one might expect a term proportional
to $\Lambda_0$. However such a term is ruled out by the chiral symmetry of the
massless theory. If $m_0$ is set equal to zero, then the original theory is
symmetric under chiral transformations of the form $\psi \to
\exp(i\omega\gamma_5) \psi$. Changing the cutoff does not affect this symmetry
and therefore any new interaction that violates chiral symmetry must vanish if
$m_0$ vanishes. Thus upon calculating the coefficient of the new $\psib\psi$
interaction one finds that it is proportional to $m_0$ rather than $\Lambda_0$;
that is, the mass renormalization is logarithmically divergent rather than
linearly divergent. Note by way of contrast that chiral symmetry is explicitly
broken in Wilson's implementation of fermions on a lattice, and consequently the
mass renormalization in such a theory is proportional to $\Lambda_0$ rather than
$m_0$.
 }
Thus the theory with Lagrangian
 \beginEq
\Lag_\Lambda = \psib(i\partial\cdot\gamma - e_\Lambda A\cdot\gamma - m_\Lambda)
\psi - \half {(F^{\mu\nu})^2} ,
 \endEq
cutoff $\Lambda$, and coupling parameters
 \beginEqarray
e_\Lambda & = & e_0 (1 + c_0(\Lambda/\Lambda_0)) \\
m_\Lambda & = & m_0 (1 + \tilde{c}_0(\Lambda/\Lambda_0))
 \endEqarray
gives the same results as the original theory with cutoff $\Lambda_0$ (up to
corrections of $\O(1/\Lambda^2)$).

With this result we see that a change in the cutoff can be compensated by
changing the bare coupling and mass in the Lagrangian in such a way that the
low-energy physics of the theory is unaffected. This is the classical result of
renormalization theory. 

 \begin{Exercise} 
Sketch out arguments for the validity of this result to
two-loop order by examining some simple process like electron-electron
scattering. Divide each loop in a diagram into high-energy and low-energy parts,
with $\Lambda$ as the dividing line. This means that each two-loop
diagram will be divided into four contributions depending upon loop energies:
high-high, high-low, low-high, and low-low. The low-low contribution is still
present with the new cutoff. The low-high and high-low contributions are removed
by the cutoff, but these are either local in character, down by $1/\Lambda$, 
or are automatically simulated in the new theory by diagrams in which the
high-energy loop is replaced by one of the new interaction introduced above
to correct one-loop results (Eqs.~\Eqnum{de} and~\Eqnum{dm}). The part of the
low-high and high-low contributions that is local can be treated with the
high-high contribution. The high-high contribution must be completely local and
can be simulated by a local interaction in the Lagrangian. Again by
power-counting, only the electron-photon vertex and the electron mass are
appreciably changed by the cutoff, and thus these new local interactions are
taken care of by introducing  corrections of relative order $e_0^4$ into
$e_\Lambda$ and $m_\Lambda$.
 \end{Exercise}

 \begin{Exercise} 
Parameters $e_\Lambda$ and $m_\Lambda$ vary as the cutoff $\Lambda$ is varied.
The couplings are said to ``run'' as more or less of the state space is included
in the cut-off theory. Show that these couplings satisfy ``evolution equations''
of the form:
 \beginEqarray
\Lambda \frac{d e_\Lambda}{d\Lambda} & = & \beta(e_\Lambda) \\
\Lambda \frac{d m_\Lambda}{d\Lambda} & = & m_\Lambda\gamma_m(e_\Lambda).
 \endEqarray
(In these equations we assume that $m_\Lambda$ is negligible compared
with $\Lambda$; more generally $\beta$ and $\gamma_m$ depend also on the ratio
$m_\Lambda/\Lambda$.) Compute $\beta$ and $\gamma_m$ to leading order in
$e_\Lambda$ for QED.

The bare parameters $e_\Lambda$ and $m_\Lambda$ can be thought of as the
effective charge and mass of an electron at energy-momentum scales of
$\O(\Lambda)$. This is particularly relevant in analyzing a
process that is characterized by only a single scale, say $Q$. To compute the
amplitude for such a process in a cut-off theory one must take $\Lambda$ much
larger than $Q$. However the main effect of vertex and self-energy corrections
to the amplitude is to replace $e_\Lambda$ and $m_\Lambda$ by $e_Q$ and $m_Q$
everywhere in the amplitude. Thus such amplitudes are most naturally expressed in
terms of the bare parameters for the theory with cutoff $\Lambda\equal Q$. This
result is not surprising insofar as physical amplitudes are independent of the
actual value of the cutoff used. The natural way to express this independence is
to calculate with $\Lambda\!\gg\!Q$ but then to reexpress the result in terms of
the running parameters at scale $Q$. In this way one removes all explicit
reference to the actual cutoff, and in particular one removes large logarithms
of $\Lambda/Q$ that otherwise tend to spoil the convergence of perturbation
theory. This procedure, while useful in QED, has proven essential in perturbative
QCD where the convergence of perturbation theory is marginal at best.
 \end{Exercise}

\subsection{Beyond Renormalizability}

It is clear from our analysis in the last section that the cut-off theory is
accurate only up to corrections of $\O(p^2/\Lambda^2)$ where $p$ is typical of the
external momenta. In practice such errors may be negligible, but if one wishes
to remove them there are two options. One is the traditional option of taking
$\Lambda$ to infinity. The other is to keep $\Lambda$ finite but to add
further corrections to the Lagrangian. This second choice is far more
informative and useful.

To illustrate the procedure consider again the $k\greater\Lambda$ part of the
amplitude for an electron scattering on an external field as in \Eq{ver}. In our
earlier analysis we neglected external momenta and masses relative to the loop
momentum. We can correct this approximation by making a Taylor expansion of the
amplitude in powers of $p/\Lambda$, $p^\prime/\Lambda$, and $m_0/\Lambda$ to
obtain terms of the form
 \beginEqarray
T(k>\Lambda) & = & -i e_0 c_0 \, \ub\,A_\ext\cdot\gamma\,u \nl
 & & - \frac{i e_0 m_0 c_1}{\Lambda^2} \,\ub\,A_\ext^\mu \sigma_{\mu\nu} 
     (p-p^\prime)^\nu \,u \nl
 & & - \frac{i e_0 c_2}{\Lambda^2}\, (p-p^\prime)^2\, 
     \ub\,A_\ext\cdot\gamma\,u 
 + \ldots ,
 \endEqarray
where coefficients $c_0$, $c_1$\ldots are all dimensionless, and where the
structure of the amplitude is constrained by the need for current
conservation and for chiral invariance in the limit $m_0\equal 0$. The effects of
all of these terms can be simulated by adding new local interactions to the
Lagrangian. The first term was handled in the previous section by simply
replacing the bare charge $e_0$ by $e_\Lambda$; the only difference now is that
contributions to $c_0$ of $\O(m_0^2/\Lambda^2)$ must be retained. The remaining
terms in $T(k\greater\Lambda)$ require the introduction of new types of
interaction:
 \beginEq
\delta\Lag_2^a = \frac{e_0 m_0 c_1}{\Lambda^2} \, 
     \psib\,F^{\mu\nu}\sigma_{\mu\nu}\,\psi 
 + \frac{e_0 c_2}{\Lambda^2} \,\psib\,i\partial_\mu F^{\mu\nu} \gamma_\nu\,\psi.
 \endEq
These interactions are designed by including a field for each external particle
in $T(k\greater\Lambda)$, and a derivative for each power of an external
momentum. By augmenting the Lagrangian with such terms we can systematically
remove all $\O(p^2/\Lambda^2)$ errors from the cut-off theory.  Of course such
errors arise in processes other than electron scattering off a field, and further
terms must be added to the Lagrangian to compensate for these. For example, the
$p^2/\Lambda^2$ contribution coming from $k\greater\Lambda$ in electron-electron
scattering (\Eq{eeee}) is compensated by interactions like
 \beginEq
\delta\Lag_2^b = \frac{d}{\Lambda^2}\,(\psib\gamma_\mu\psi)^2 .
 \endEq
Luckily power-counting and dimensional analysis tell us that only a few
processes are affected by the new cutoff to this order, and therefore only a
finite number of terms need be added to the Lagrangian to remove all errors of 
$\O(p^2/\Lambda^2)$ in all processes. 

It seems remarkable that the $p^2/\Lambda^2$ errors for an infinity of
processes can removed from the cut-off theory by adding a finite (even small)
number of new interactions to the Lagrangian. However, one can show, even without
examining particular processes, that there is only a finite
number of possible new interactions that might have been relevant to this order.
Interactions that simulate $k\greater\Lambda$ physics must be local---i.e.,
polynomial in the fields, and derivatives of the fields---and they must have the
same symmetries as the underlying theory. In QED, these symmetries include Lorentz
invariance, gauge invariance, parity conservation, and so on.\footnote{
 If the regulator breaks one of the symmetries of the theory then interactions
that break the symmetry will also arise. These interactions serve to cancel
the symmetry-breaking effects of the regulator. In lattice gauge theory, for
example, the lattice regulator breaks the Lorentz symmetry and as a consequence
interactions in such a theory can be Lorentz {\em non}-invariant.
 }
In addition the interaction terms must have (energy) dimension four; thus an
interaction operator of dimension $n+4$ must have a coefficient of
$\O(1/\Lambda^{n})$ or smaller, $\Lambda$ being the smallest scale in the
$k\greater\Lambda$ part of the theory. Dimension-six operators are obviously
important in $\O(1/\Lambda^2)$, but operators of higher dimension are
suppressed by additional powers of $1/\Lambda$ and so are irrelevant to this
order. There are very few operators of dimension six or less that are Lorentz
and gauge invariant, and that can be constructed from polynomials of $\psi$
(dimension $3/2$), $A_\mu$ (dimension $1$), and $\partial_\mu$ (dimension $1$).
And these few are the only ones needed to correct the Lagrangian of the cut-off
theory through order $p^2/\Lambda^2$.

 \begin{Exercise}
It is critical to our discussion that an operator of dimension $n+4$
in the Lagrangian only affect results in order $(p/\Lambda)^n$ (or less). So
introducing, for example, a dimension-eight operator should not affect the
predictions of the theory through order $(p/\Lambda)^2$. This is obviously the
case at tree level in perturbation theory, since the coefficient of the
dimension-eight operator has a factor $1/\Lambda^4$ that must appear in the
final result for any tree diagram involving the interaction. Such a contribution
will then be suppressed by a factor $(p/\Lambda)^4$. However in one-loop order
(and beyond) loop integrations can supply powers of $\Lambda$ that cancel the
powers of $1/\Lambda$ explicit in the interaction, resulting in contributions
from the new interaction that are not negligible in second order. Show that
these new contributions can be cancelled by appropriate shifts in the
coefficients of the lower dimension operators in the Lagrangian. Thus the
dimension-eight operator has no net effect on the results of the theory through
order $p^2/\Lambda^2$.
 \end{Exercise}

This procedure can obviously be extended so as to correct the theory to any
order in $p/\Lambda$, but the price of this improved accuracy is a more
complicated Lagrangian. So why bother with cut-off field theory? Certainly in
perturbation theory it seems desirable that the number of interactions be kept as
few as possible  so as to avoid an explosion in the number of diagrams. On the
other hand the theory with no cutoff or with $\Lambda\!\to\!\infty$ is
ill-defined. Furthermore in practice it is essential to reduce the number of
quantum degrees of freedom before one is able to solve a quantum field theory. A
cutoff does just this. While  cutoffs tend to appear only in intermediate steps of
perturbative analyses, they are of central importance in most nonperturbative
analyses. In numerical treatments, using lattices for example, the number of
degrees of freedom must be finite since computers are finite. 

The practical consequences of our analysis are important, but the most striking
implication is that nonrenormalizable interactions make sense in
the context of a cut-off field theory. Insofar as experiments can probe only a
limited range of energies it is natural to analyze experimental results using
cut-off field theories. Since nonrenormalizable interactions occur naturally
in such theories it is probably wrong to ignore them in constructing
theoretical models for currently accessible physics.

 \begin{Exercise}
The idea behind renormalization is rather generally applicable in quantum
mechanics. Suppose we have a problem in which the states of most interest all
have low energies, but couple through the Hamiltonian to states with much higher
energies. To simplify the problem we want to truncate the state space, excluding
all states but the (low-energy) ones of interest. Suppose we define a projection
operator $\P$ that projects onto this low-energy subspace; its complement is
$\Q\equiv 1-\P$. Then show that the full Hamiltonian $H$ for the theory can be
replaced by an effective Hamiltonian that acts only on the low-energy subspace:
 \beginEq
H_\eff(E) = H_{\P\P} + H_{\P\Q} \frac{1}{E-H_{\Q\Q}} H_{\Q\P}
 \endEq
where $H_{\P\Q}\equiv\P H \Q$\ldots. In particular if $\ket{E}$ is  an
eigenstate of the full Hamiltonian then its projection onto the low-energy
subspace $\ket{E}_\P \equiv \P \ket{E}$ satisfies
 \beginEq
H_\eff(E) \ket{E}_\P = E \ket{E}_\P .
 \endEq
Note that since the energies $E$ in the $\P$-space are much smaller than the
energies in the $\Q$-space, the second term in $\H_\eff$ can be expanded in a
powers of $E/H_{\Q\Q}$. Thus the new interactions are polynomial in $E$ and
therefore ``local'' in time. In practice only a few terms might be needed in this
expansion. Such truncations are made all the time in atomic physics. For example,
if one is doing radio-frequency studies of the hyperfine structure of the ground
state of hydrogen one usually wants to forget about all the radial excitations of
the atom (i.e., optical frequencies). 
 \end{Exercise}

 \subsection{Structure and Interpretation of Cut-off Field Theories}

In the preceding sections we illustrated the nature of a cut-off field theory
using QED perturbation theory. It should be emphasized that most of what we
discovered is not tied to QED or to perturbation theory. The central
notion, that the effects of high-energy states on low-energy processes can be
accounted for through the introduction of local interactions, follows from the
uncertainty principle and so is quite general. We can summarize the general
ideas underlying cut-off field theories as follows:
 \begin{itemize}
 \item A finite cutoff $\Lambda$ can be introduced into any field theory for
the purposes of discarding high-energy states from the theory. The cut-off theory
can then be used for processes involving momenta $p$ much less than $\Lambda$.

 \item The effects of the discarded states can be retained in the theory by
adjusting the existing couplings in the Lagrangian, and by adding new, local,
nonrenormalizable interactions. These interactions are polynomial in the fields
and derivatives of the fields. The nonrenormalizable couplings do not result in
intractable infinities since the theory has a finite cutoff.

 \item Only a finite number of interactions is needed when working to a
particular order in $p/\Lambda$, where $p$ is a typical momentum in whatever
process is under study. The cutoff usually sets the scale of the coefficient of
an interaction operator: the coefficient of an operator with (energy) dimension
$n+4$ is $\O(1/\Lambda^n)$, unless symmetries or approximate symmetries of the
theory further suppress the interaction. Therefore one must include all
interactions involving operators with dimension $n+4$ or less to achieve accuracy
through order $(p/\Lambda)^n$.

 \item Conversely an operator of dimension $n+4$ can only affect results at
order $(p/\Lambda)^n$ or higher, and so may be dropped from the theory if such
accuracy is unnecessary.
 \end{itemize}

 Note that the interactions in a cut-off field theory are
completely specified by the requirement of locality, by the symmetries of the
theory and regulator, and by the accuracy desired. The structure of the
operators introduced into the Lagrangian does not depend upon the detailed
dynamics of the high-energy states being discarded. It is only the numerical
values of the coefficients of these operators that depend upon the high-energy
dynamics. Thus while the high-energy states do have a strong effect on low-energy
processes we need know very little about the high-energy sector in order to
compute low-energy properties of the theory. The coupling constants of the cut-off
theory---$e_\Lambda$, $m_\Lambda$, $c_1$, $c_2$, $d$\ldots{}in the
previous sections---completely characterize the behavior of the discarded
high-energy states for the purposes of low-energy analyses. In cases where we
understand the dynamics of the discarded states, as in our analysis of QED, we
can compute these coupling constants. In other situations we are compelled to
measure them experimentally.

The expansion of a cut-off Lagrangian in powers of $1/\Lambda$ is somewhat
analogous to multipole expansions used in classical field theory. For example,
the detailed charge distribution of a nucleus is of little importance to an
electron in an atomic orbital. To compute the long range electrostatic field of
the nucleus one need only know the charge of the nucleus, and, depending upon
the level of accuracy desired, perhaps its dipole and quadrupole moments. Again,
for all practical purposes the effects of short-range structure on long-range
behavior can be expressed in terms of a finite number of numbers, the multipole
moments, characterizing the short-range structure.

Armed with this new understanding of field theory, it is time to reexamine
traditional theories in an effort to better understand why they are the way
they are and how they might be changed to accommodate future experiments.

 \section{Applications}

 \subsection{Why is QED renormalizable?}
Having argued that nonrenormalizable interactions are admissible one has to
wonder why QED is renormalizable after all. What do we learn from the fact of
its renormalizability? Our new attitude towards renormalization suggests that
the key issue in addressing a theory like QED is not whether or not it is
renormalizable, but rather how renormalizable it is---i.e., how large are the
nonrenormalizable interactions in the theory.

It is quite likely that QED is a low-energy approximation to some complicated
high-energy supertheory (a string theory?). Consequently there will be some large
energy scale beyond which QED dynamics are insufficient to describe nature, where
the supertheory will become necessary. Since we have yet to figure out what the
supertheory is, it is natural in this scenario that we introduce a cutoff
$\Lambda$ equal to this energy scale so as to exclude the unknown physics. The
supertheory still affects low-energy phenomena, but it does so only through the
values of the coupling constants that appear in the cut-off Lagrangian:
 \beginEqarray
\Lag_\Lambda &=& \psib(i\partial\cdot\gamma - e_\Lambda A\cdot\gamma - m_\Lambda)
\psi - \half {(F^{\mu\nu})^2}  + \nl
&& + \frac{e_\Lambda m_\Lambda c_1}{\Lambda^2} \, 
     \psib\,F^{\mu\nu}\sigma_{\mu\nu}\,\psi 
 + \frac{e_\Lambda c_2}{\Lambda^2} \,\psib\,i\partial_\mu F^{\mu\nu}
\gamma_\nu\,\psi  + \frac{d}{\Lambda^2}\,(\psib\gamma_\mu\psi)^2  + \ldots.
 \endEqarray
We cannot calculate the coupling constants in this Lagrangian until we discover
and solve the supertheory; the couplings must be measured. The nonrenormalizable
interactions are certainly present, but if $\Lambda$ is large their affect 
on current physics will be very small, down by $(p/\Lambda)^2$ or more. The
fact that we haven't needed such terms to account for the data tells us that
$\Lambda$ is indeed large. This is the key to the significance of
renormalizability and its origins: very low-energy approximations to
{\em arbitrary} high-energy dynamics can be formulated in terms of 
renormalizable field theories since nonrenormalizable interactions would be
suppressed in their effects by powers of $p/\Lambda$ and therefore would be
irrelevant for $p\!\ll\!\Lambda$. 
This analysis also tells us how to look for
low-energy evidence of new high-energy dynamics: look for 
(small) effects caused by the leading nonrenormalizable interactions in the
theory. Indeed we can estimate the energy scale $\Lambda$ at which new physics
must appear simply by measuring the strength of the nonrenormalizable
interactions in a theory. The low-energy theory must fail and be replaced at
energies of order $\Lambda$. Processes with $p\!\sim\!\Lambda$ start to probe the
detailed structure of the supertheory, and can no longer be described by the
low-energy theory; the expansion of the Lagrangian in powers of $1/\Lambda$ no
longer converges. 

As we noted, the successes of renormalizable QED imply that the energy threshold
for new physics in electrodynamics is rather high. For example, the
$\psib\sigma\cdot F \psi$ interaction in $\Lag_\Lambda$ would shift the
$g$-factor in the electron's magnetic moment by an amount of order
$(m_e/\Lambda)^2$. So the fact that renormalizable QED accounts for the
$g$-factor to almost 12 digits implies that
 \beginEq
\frac{m_e^2}{\Lambda^2} < 10^{-12}
 \endEq
from which we can conclude that $\Lambda$ is probably larger than a TeV.

\subsection{Strong Interactions: Pions or Quarks?}

QED is spectacularly successful in explaining the magnetic moment of the
electron. It's failure to explain the magnetic moment of the proton is equally
spectacular: $g_p$ is roughly three times larger than predicted by QED when the
proton is treated as an elementary, point-like particle. Nonrenormalizable terms
in $\Lag_\Lambda$ make contributions of order unity,
 \beginEq
\frac{m_p^2}{\Lambda^2} \sim 1 ,
 \endEq
indicating that the cutoff in proton QED must be of order the proton's mass
$m_p$. Thus  QED with an elementary proton must fail around \GeV{1}, to be
replaced by some other more fundamental theory. That new theory is quantum
chromodynamics, of course, and the large shift in the proton's magnetic moment
is a consequence of its being a composite state built of quarks.  This example
illustrates how the measured strength of nonrenormalizable interactions can be
used to set the energy scale for the onset of new physics. Also it strongly
suggests that one ought to use QCD in analyzing strong interaction physics above
a GeV.

This result also suggests that one can and ought to treat the proton as a
point-like particle in analyses of sub-GeV physics. In the hydrogen atom, for
example, typical energies are of order a few eV. At such energies the proton's
electromagnetic interactions are most efficiently described by a cut-off QED (or
nonrelativistic QED) for an elementary proton. The cutoff $\Lambda$ should be set
below a GeV, and the anomalous magnetic moment, charge radius and other
properties of the extended proton should be simulated by nonrenormalizable
interactions of the sort we have discussed. This theory can be made arbitrarily
precise by adding a sufficient number of interactions---an expansion in
$1/\Lambda$---and by working to sufficiently high order in perturbation
theory---an expansion in $\alpha$. With the cutoff in place, the
nonrenormalizable interactions cause no particular problems in perturbation
theory.

In the same spirit one expects very low-energy strong interactions to be
explicable in terms of a theory of point-like mesons and baryons. Indeed these
ideas account for the great success of PCAC and current algebra in describing
pion-pion and pion-nucleon scattering at threshold. A cut-off theory of
elementary pions, protons and other hadrons must work above threshold as well,
but will ultimately fail somewhere around a GeV. It is still an open question as
to just where the cutoff lies. In particular it is unclear
whether or not nuclear physics can still be efficiently described by pion-nucleon
theories at energies of a few hundred MeV, or whether the quark structure is
already important at such energies. To evaluate the utility of the pion-nucleon
theory one must treat it as a cut-off field theory, taking care to make consistent
use of the cutoff throughout, and systematically enumerating the possible
interactions, determining their couplings from experimental data. It also seems
likely that some sort of (systematic) nonperturbative approach to the problem is
essential. This is clearly a crucial issue for nuclear physics since a theory of
point-like hadrons, if it works, should be far simpler to use than a theory of
quarks and gluons.

\subsection{Weak Interactions: Renormalization, Naturalness and the SSC}

The Standard Model of strong, electromagnetic and weak interactions has been
enormously successful in accounting for experimental data up to energies of
order \GeV{100} or more. Still there remain several unanswered questions, and
the most pressing of these concerns the origins of the masses for the $W$ and
$Z$ bosons. A whole range of possible mechanisms has been suggested---the Higgs
mechanism, technicolor, a supersymmetric Higgs particle, composite vector
bosons---and extensive experimental searches have been conducted. Yet we still do
not have a definitive answer. What will it take to unravel this puzzle?

In fact new physics, connected with the $Z$ mass, must appear at energies of order
a few TeV or lower. Current data indicates that the vector bosons are massive
and that their interactions are those of nonabelian gauge bosons, at least
insofar as the quarks are concerned. This suggests that a minimal Lagrangian for
the vector bosons that accounts for current data would be the standard
Yang-Mills Lagrangian for nonabelian gauge fields plus simple mass terms for the
$W$'s and $Z$.  Traditionally such a theory is rejected immediately since the
mass terms spoil gauge invariance and ruin renormalizability. However from our
new perspective, the appearance of nonrenormalizable terms in the minimal theory
poses no problems; it indicates that this theory is necessarily a cut-off field
theory with a {\em finite\/} cutoff. Thus the theory with minimal particle
content becomes inadequate at a finite energy of order the cutoff, and new
physics is inevitable. By examining the nonrenormalizable interactions one finds
that the value of the cutoff is set by the vector-boson mass $M$:\footnote
 {It is the longitudinal degrees-of-freedom in a massive Yang-Mills
theory that spoil the renormalization. The longitudinal part of the action can be
isolated through gauge transformations and shown to be equivalent to a nonlinear
sigma model, which is nonrenormalizable. To see this simply, one can examine the
standard theory with a Higgs particle. One arranges the scalar couplings of this
theory in such a way that the mass of the Higgs becomes infinite, while keeping
the vector-boson masses constant. What remains, once the Higgs particle is removed
in this fashion, is just a theory of massive gauge bosons. The part of the
Lagrangian dealing with the Higgs particle becomes a nonlinear sigma model in
this limit, with the magnitude of the complex scalar field frozen at its vacuum
expectation value $v\approx\mbox{\GeV{200}}$. The sigma model is
nonrenormalizable but makes sense as a cut-off theory provided the cutoff is less
than $\O(4\pi v)$, which is equivalent to the limit given in the text.
 }
 \beginEq
\Lambda \sim \frac{\sqrt{4\pi}M}{\sqrt{\alpha}}
 \labelEq{wkcut}
 \endEq
This cutoff is of order a few TeV, and it represents the threshold energy for
new physics. 

This analysis is the basis for one of the most important scientific arguments for
building an accelerator, like the SSC, that can probe few-TeV physics; the
secret behind the $W$ and $Z$ masses almost certainly lies in this energy range.
One potential problem with this argument is the possibility that the Higgs
particle exists and has a mass well below a few TeV, in which case the
interesting physics is at too low an energy for an SSC. In fact the argument
survives because a theory with a low-mass Higgs particle is ``unnatural'' unless
there is new structure, again at energies of order a few TeV. If we regard the
standard model with a Higgs field as a cut-off field theory, approximating some
more complex high-energy theory, then we expect that the scale of dimensionful
couplings in the cut-off Lagrangian is set by the cutoff $\Lambda$. In particular
the bare mass of the scalar boson is of order the cutoff, and thus the
renormalized mass of the Higgs particle ought to be of this order as well.
Since this theory is technically renormalizable, one can make the Higgs mass much
smaller than the cutoff by fine tuning the bare mass to cancel almost exactly
the large mass generated by quantum fluctuations; but such tuning is highly
contrived, and as such is unlikely to occur in the low-energy approximation to any
more complex theory. Thus the Higgs theory is sensible only when the cutoff
$\Lambda$ is finite. To estimate how large a natural cutoff might be, we compare
the bare mass with the mass renormalization due to quantum fluctuations. Barring
(unlikely) accidental cancellations, one expects the physical and bare masses to
have roughly the same order of magnitude; that is, we expect
 \beginEq
m_H^2 \equiv  \mu^2 + \delta m^2 
 \sim  |\mu^2|,
 \endEq
where $\delta m^2$ is the mass renormalization, and $m_H^2$ and $\mu^2$ are the
squares of the physical and bare masses respectively. 
The mass renormalization is quadratically sensitive to the cutoff,
 \beginEq
\delta m^2 \sim \lambda\Lambda^2
 \endEq 
where $\lambda$ is the $\phi^4$ coupling constant. Therefore the bare and
physical masses can be comparable  only if the cutoff is less than
 \beginEq
\Lambda^2 \sim \frac{|\mu^2|}{{\lambda}},
 \endEq
which turns out to be the same limit we obtained above for the
nonrenormalizable massive Yang-Mills theory (\Eq{wkcut}).\footnote{
 Recall that the complex scalar field generates a mass for the gauge bosons by
acquiring a vacuum expectation value. Replacing $i\partial_\mu$ by
$i\partial_\mu-g A_\mu$ in the Lagrangian for a free scalar boson leads to an
interaction term $g^2 A^2 \phi^2/2$. This becomes a mass term for the $A$ field
if the $\phi$ field has nonzero vacuum expectation value $v$, the mass
being $gv$. Thus $v$ is of order $M/\sqrt{\alpha}$. This $v$ arises from the
competition between the bare scalar mass term and the $\phi^4$ interaction, with
the result that $v^2$ is also of order $|\mu^2|/\lambda$, which in turn is of
order the cutoff squared in a natural theory. Combining these two expressions for
$v$ gives  the limit  in \Eq{wkcut}.
 }
Therefore new physics is expected in the few-TeV region whether or not there is
a Higgs particle at lower energies.

The Yang-Mills theory coupled to a Higgs doublet gained widespread acceptance
because it was renormalizable. However, our new understanding of cut-off field
theories suggests that naturalness, rather than renormalizability, is the key
property of a physical theory. And from this perspective, the renormalizable
theory with a Higgs doublet is neither better nor worse than the
nonrenormalizable massive Yang-Mills theory. One might even argue that
the latter model is the more attractive given that it is minimal. Both
theories are predictive in the \GeV{100} region and below; neither theory can
survive without modification much beyond a few TeV.

Our analysis of the Higgs sector of the standard weak interaction theory
illustrates a general feature of physically relevant scalar field theories: the
renormalized mass of the field is most likely as large as the bare mass, and
both are at least as large as the mass renormalization due to quantum
fluctuations. In strongly interacting theories this means that the renormalized
mass is of order the cutoff; scalars with masses small compared to $\Lambda$ are
unnatural. This result should be contrasted with the situation for spin-1/2
particles and for gauge bosons. In the case of vector bosons like the photon,
the bare mass in the Lagrangian's mass term, $M_\Lambda^2 A^2/2$,
would indeed be of order the cutoff were it not for gauge invariance, which
requires $M_\Lambda \equal 0$. Similarly for fermions like the electron, chiral
symmetry implies that all mass renormalizations must vanish if the bare mass
does. Consequently one has $\delta m_e\!\sim\! m_e$ even though the cutoff is
much larger. In general one expects low-mass particles in a theory only when
there is a symmetry, like gauge invariance or chiral symmetry, that protects the
low mass. No such symmetry exists for scalar bosons, unless they are tied to
fermionic partners via a supersymmetry. This may explain why all existing
experimental data in high-energy physics can be understood in terms of elementary
particles that are either spin-1/2 fermions or spin-1 gauge bosons.

\subsection{Nonrelativistic Field Theories}

Nonrelativistic systems, such as positronium or the $\Upsilon$ and $\psi$ mesons,
play an important role in several areas of elementary particle physics. Being
nonrelativistic these systems are generally weakly coupled, and as a result
typically involve only a single channel. Thus, for example, the $\Upsilon$ is
predominantly a bound state of a $b$ quark and antiquark. It has some probability
for being in a state comprised of a $b\overline{b}$ pair and a gluon, or a
$b\overline{b}$ pair and a $u\overline{u}$ pair, etc., but the probabilities are
small and therefore these channels have only a small effect on the gross physics
of the meson. This is an enormous simplification relative to relativistic systems
where many channels may be important, and it is this that accounts for the
prominence of such systems in fundamental studies of electromagnetic and strong
interactions. Nevertheless there are significant technical problems connected
with the study of nonrelativistic systems. Central among these is the problem of
too many energy scales. Typically a nonrelativistic system has three important
energy scales: the masses $m$ of the particles involved, their three-momenta
$p\!\sim\! m v$, and their kinetic energies $KE\!\sim\! m v^2$. These scales are
widely different in a nonrelativistic system since $v\!\ll\! 1$ (where the speed
of light  $c = 1$), and this greatly complicates any analysis of such a
system. In this section we will see how renormalization can be used as a tool to
dramatically simplify studies of this sort.  

One can appreciate the problems that arise when analyzing these
systems by considering a lattice simulation of the $\Upsilon$ meson. The
space-time grid used in such a simulation must accommodate wavelengths covering all
of the of scales in the meson, ranging from $1/mv^2$ down to $1/m$. Given that
$v^2\!\sim\! 0.1$ in the $\Upsilon$, one might easily need a lattice as large as
100 sites on side to do a good job. Such a lattice could be three times larger
than the largest wavelength, with a grid spacing three times smaller than the
smallest wavelength, thereby limiting the errors caused by the grid. This is a
fantastically large lattice by contemporary standards and is quite impractical.

The range of length scales also causes problems in analytic analyses. As an
example, consider a traditional Bethe-Salpeter calculation of the energy levels
of positronium. The potential in the Bethe-Salpeter equation is given by a sum of
two-particle irreducible Feynman amplitudes. One generally solves the problem
for some approximate potential and then incorporates corrections using
time-independent perturbation theory. Unfortunately, perturbation theory for a
bound state is far more complicated than perturbation theory for, say, the
electron's $g$-factor. In the latter case a diagram with three photons
contributes only in order $\alpha^3$.
% \Figspace{ps1}{1.57in by 0.76in}{A two-loop kernel contributing to the
%          Bethe-Salpeter potential for positronium.}
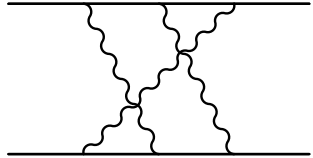
\begin{figure}
\begin{center}
\begin{fmfgraph*}(40,20)
\fmfstraight
\fmftopn{t}{5}
\fmfbottomn{b}{5}
\fmfcyclen{plain}{t}{5}
\fmfcyclen{plain}{b}{5}
\fmf{boson}{b2,t4}
\fmf{boson}{b3,t2}
\fmf{boson}{b4,t3}
\end{fmfgraph*}
\end{center}
\caption{\label{fig:ps1}A two-loop kernel contributing to the
          Bethe-Salpeter potential for positronium.}
\end{figure}
In positronium a kernel involving the
exchange of three photons (e.g., \Fig{ps1}) can also contribute to order
$\alpha^3$, but the same kernel will contribute to all higher orders as
well:\footnote{The situation is actually even worse. The contribution from a
particular kernel is highly dependent upon gauge. For example, the coefficients
$a_0$ and $a_1$ vanish in Coulomb gauge but not in Feynman gauge. The Feynman
gauge result is roughly $10^4$ times larger, and it is spurious: the large
contribution comes from unphysical retardation effects in the Coulomb
interaction that cancel when an infinite number of other diagrams is included.
The Coulomb interaction is instantaneous in Coulomb gauge and so this gauge
generally does a better job describing the fields created by slowly moving
charged particles. On the other hand contributions coming from relativistic
momenta are more naturally handled in a covariant gauge like Feynman gauge; in
particular renormalization is far simpler in Feynman gauge than it is in Coulomb
gauge. Unfortunately most Bethe-Salpeter kernels have contributions coming from
both nonrelativistic and relativistic momenta, and so there is no optimal choice
of gauge. This is again a problem due to the multiple scales in the system.}
 \beginEq
\avg{ K_3 } = \alpha^3 m \left\{ a_0 + a_1\alpha + a_2\alpha^2 +\ldots \right\}.
 \labelEq{ps2}
 \endEq
So in the bound state
calculation there is no simple correlation between the importance of an amplitude
and the number of photons in it. Such behavior is at the root of the complexities
in high-precision analyses of positronium or other QED bound states, and it is a
direct consequence of the multiple scales in the problem. Any expectation value
like that in \Eq{ps2} will be some complicated function of ratios of the three
scales in the atom:
 \beginEq
\avg{K_3} = \alpha^3 m \; f\!\left( {\avg{p}}/{m},{\avg{KE}}/{m} \right).
 \endEq
Since $\avg{p}/m\sim\alpha$ and $\avg{KE}/m\sim\alpha^2$, a Taylor expansion of
$f$ in powers of these ratios generates an infinite series of contributions just
as in \Eq{ps2}. Similar series do not occur in the $g$-factor
calculation because there is but one scale in that problem, the mass of the
electron.

Traditional methods for analyzing either of these systems fail to take advantage
of the nonrelativistic character of the systems. One way to capitalize on this
feature is to introduce a cutoff  of order the constituent's mass or smaller into
the field theory. The cutoff here can be thought of as the boundary between
relativistic and nonrelativistic physics. Since the gross dynamics in the
problems of interest is nonrelativistic, relativistic physics is well simulated
by local interactions in the cut-off Lagrangian. 

The utility of such a cutoff is greatly enhanced if one also transforms the Dirac
field so as to decouple its upper components from its lower components. This is
called the Foldy-Wouthuysen transformation, and it transforms the Dirac
Lagrangian into a nonrelativistic Lagrangian. In QED one obtains
 \beginEqarray
\overline{\Psi}(iD\cdot\gamma - m) \Psi 
 & \to & \psid  iD_0 \psi + \psid \frac{\Dv^2}{2m} \psi 
    - \psid\frac{\Dv^4}{8m^3}\psi      \nl
 && - \frac{e}{2m}\psid\sigmav\cdot\Bv\psi
    - \frac{e}{8m^2}\psid\delv\cdot\Ev\psi+\cdots
 \endEqarray
where $D_\mu = \partial_\mu + i e A_\mu$ is the gauge-covariant derivative,
$\Ev$ and $\Bv$ are the electric and magnetic fields, and $\psi$ is a
two-component Pauli spinor representing the electron part, or upper components, of
the original Dirac field. The lower components of the
Dirac field lead to analogous terms that specify the electromagnetic interactions
of positrons. The Foldy-Wouthuysen transformation generates an infinite expansion
of the action in powers of $1/m$. As an ordinary $\Lambda\!\to\!\infty$ field
theory this expansion is a disaster; the renormalizability of the theory is
completely disguised, requiring a delicate conspiracy involving terms of all
orders in $1/m$. However, setting $\Lambda\!\sim\! m$ implies that the
Foldy-Wouthuysen expansion is an expansion in $1/\Lambda$, and from our general
rules for cut-off theories we know that only a finite number of terms need be
retained in the expansion if we want to work to some finite order in $p/\Lambda
\sim p/m \sim v$. Thus, to study positronium through order $v^2\!\sim\! \alpha^2$,
we can replace QED by a nonrelativistic QED (NRQED) with the Lagrangian
 \beginEqarray
\Lag_{NRQED} & = &  - \frac{(F^{\mu\nu})^2}{2} +
  \psid \left\{ i\partial_t - e_0 A_0 + \frac{\Dv^2}{2m_0} \right. \nl
&& - c_1\frac{e_0}{2m_0}\sigmav\cdot\Bv - c_2\frac{e_0}{8m_0^2}\delv\cdot\Ev \nl
&& - c_3\left.\left( \frac{ie_0}{8m_0^2} \sigmav\cdot\delv\times\Ev 
     - \frac{ie_0}{4m_0^2} \sigmav\cdot\Ev\times\delv \right) \right\} \psi \nl
&& + \frac{d_1}{m_0^2}(\psid\psi)^2 + \frac{d_2}{m_0^2} (\psid\sigmav\psi)^2 \nl
&& + \mbox{positron and positron-electron terms}. 
 \endEqarray
The coupling constants $e_0$, $m_0$, $c_1$\ldots are all specified for a
cutoff of $\Lambda\!\sim\! m$. Renormalization theory tells us that there exists a
choice for the coupling constants in this theory such that NRQED reproduces all
of the results of QED up to corrections of order $(p/m)^3$.

NRQED is far simpler to use than the original theory when studying nonrelativistic
atoms like positronium. The analysis falls into two parts. First one must
determine the coupling constants in the NRQED Lagrangian. This is easily done by
computing simple scattering amplitudes in both QED and NRQED, and then by
adjusting the NRQED coupling constants so that the two theories agree through
some order in $\alpha$ and $v$. The coupling constants are functions of $\alpha$
and the mass of the electron; to leading order one finds that the $d_i$'s vanish
while all the $c_i$'s equal one. As the couplings contain the relativistic
physics, this part of the calculation involves only scales of order $m$; it is
similar in character to a calculation of the $g$-factor. Furthermore there is no
need to deal with complicated bound states at this stage. Having solved the
high-energy part of QED by computing $\Lag_{NRQED}$, one goes on to solve
NRQED for any nonrelativistic process or system. To study positronium one uses
the Bethe-Salpeter equation for this theory, which is just the
Schr\"{o}dinger equation, and ordinary time-independent perturbation theory. One
of the main virtues of this approach is that it builds directly on the simple
results of nonrelativistic quantum mechanics, leaving our intuition intact. Even
more important for high-precision calculations is that only two dynamical scales
remain in the problem, the momentum and the kinetic energy, and these are
easily separated on a diagram-by-diagram basis. As a result infinite series in
$\alpha$ can be avoided in calculating the contributions due to individual
diagrams, and thus it is trivial to separate, say, $\O(\alpha^6)$ contributions
from $\O(\alpha^5)$ contributions.

Obviously NRQED is useful in analyzing  electromagnetic  interactions in any
nonrelativistic system. In particular it provides an elegant framework for
incorporating relativistic effects into analyses of many-electron atoms and
solids in general.

For heavy-quark mesons one can replace QCD by a nonrelativistic theory (NRQCD)
whose Lagrangian has basically the same structure as $\Lag_{NRQED}$. Since the
coupling constants relate to physics at scales of order $m$ and above, they can
be computed perturbatively if the quark mass is large enough. The lattice used
in simulating NRQCD can be much coarser than that required for ordinary QCD since
structure at wavelengths of order $1/m$ has been removed from the theory. For
example,  $v\!\sim\!1/3$ for the $\Upsilon$ and thus the lattice spacing can
be roughly three times larger in NRQCD, leading to a lattice with $3^4$ fewer
sites. Furthermore, by decoupling the quark from the antiquark degrees of freedom
we convert the numerical problem of computing quark propagators from a
boundary-value problem for a four-spinor into an initial-value problem for a
two-spinor, resulting in very significant gains in efficiency. The move from QCD
to NRQCD makes the $\Upsilon$ one of the easiest mesons to simulate numerically.

\section{Conclusion}

In this lecture we have seen that renormalizability is not miraculous; on the
contrary, approximate renormalizability is quite natural in a theory that is a
low-energy approximation to some more complex high-energy supertheory.
Furthermore one expects symmetries like gauge invariance or chiral symmetry in
the low-energy theory; low-mass particles are unnatural without such symmetries.
In such a picture the ultraviolet cutoff, originally an artifice required to give
meaning to divergent integrals, acquires physical significance as the threshold
energy for the appearance of new physics associated with the supertheory; it is
the boundary between what we do and do not know. With the cutoff in place it is
quite natural to have small nonrenormalizable interactions in the low-energy
theory, the size of these interactions being intimately related to the energy
range over which the low-energy approximation is valid: the smaller the coupling
constants for nonrenormalizable interactions, the larger the range of validity
for the theory. Thus in studying weak, electromagnetic and strong processes we
should be on the lookout for evidence indicating such interactions since this will
give us clues as to where new physics might be found. Finally, we saw that
renormalization and cut-off Lagrangians are powerful tools that can be used to
separate energy scales in a problem, allowing us to deal with one scale at a
time. In light of these developments we need no longer apologize for
renormalization.

\end{fmffile}

\section*{Acknowledgements}
 
I thank Kent Hornbostel for his comments and suggestions concerning
this manuscript. These were most useful. This work was supported by a grant from
the National Science Foundation.

\section*{Bibliography}

Cut-off field theories have played an important role in the modern development of
the renormalization group as a tool for studying both quantum and statistical
field theories. An elementary discussion of this subject, with lots of
references to the original literature, can be found in Wilson's Nobel lecture:
 \begin{Ref} K.G.\ Wilson, {\em Rev.\ Mod.\ Phys.\/} {\bf 55}, 583(1983); see also
K.G.\ Wilson, {\em Sci.\ Am.\/} {\bf 241}, 140 (August, 1979).\end{Ref}

The effects of a finite cutoff are naturally important in numerical simulations
of lattice QCD, where it is hard enough to even get the lattice spacing small,
let alone vanishingly small. The use of nonrenormalizable interactions to
correct for a finite lattice spacing has been explored extensively; see, for
example,
 \begin{Ref} K.\ Symanzik, {\em Nucl.\ Phys.\/} {\bf B226}, 187(1983). \end{Ref}
The use of Monte Carlo techniques in renormalization-group analyses was
discussed by R.\ Gupta at this Summer School.

Another area in which cut-off Lagrangians have come to the fore is in
applications of current algebra to low-energy strong interactions. Nonlinear
chiral models provide the natural starting point for any attempt to model
such physics in terms of elementary meson and baryon fields. A good introduction
to the general ideas underlying these theories is given in Georgi's book:
 \begin{Ref} H.\ Georgi, {\em Weak Interactions and Modern Particle Theory},
             Benjamin/Cummings, Menlo Park, 1984. \end{Ref} 
Note that Georgi refers to cut-off theories as effective theories. I have
avoided this usage to prevent confusion between cut-off Lagrangians and
effective classical Lagrangians, in which all loops are incorporated into the
action. Serious attempts have been made to carry chiral theories beyond tree level
using perturbation theory; see, for example, 
 \begin{Ref} J.\ Gasser and H.\ Leutwyler, {\em Ann. Phys. (N.Y.)\/} {\bf 158},
142(1984). \end{Ref} 
Unfortunately, the convergence of perturbation theory is not always particularly
impressive here. Pion-nucleon theories are analyzed extensively in the
literature of nuclear physics. However most of these analyses are inconsistent
in the application of cutoffs and the design of the cut-off Lagrangian.

Experimental limits on possible deviations from the standard model have been
studied extensively in preparation for the SSC and other multi-TeV accelerators.
For example, a study of the limits on nonrenormalizable interactions in QED, QCD,
etc.\ can be found in the proceedings of the first Snowmass workshop on SSC
physics:
 \begin{Ref} M.\ Abolins {\em et al.\/}, in {\em Proceedings of the 1982 DPF
Summer Study on Elementary Particle Physics and Future Facilities},
edited by R.\ Donaldson {\em et al.\/}. 
 \end{Ref} 
The significance of an extra anomaly in the electron's magnetic moment,
beyond what is predicted by QED, was first discussed in
 \begin{Ref} R.\ Barbieri, L.\ Maiani and R.\ Petronzio, {\em Phys.\
Lett.\/} {\bf 96B}, 63(1980); and S.J.\ Brodsky and S.D.\ Drell, {\em Phys.\
Rev.\/} {\bf D22}, 2236(1980).
 \end{Ref}
The proceedings of the Snowmass workshops and similar workshops contain extensive
discussions of the various scenarios for the weak interactions at TeV energies.
See also
 \begin{Ref} M.S.\ Chanowitz, {\em Ann.\ Rev.\ Nucl.\ Part.\ Sci.\/} {\bf 38},
323(1988) 
 \end{Ref}
where among other things, the phenomenology of the (nonrenormalizable) massive
Yang-Mills version of weak interactions is discussed. The naturalness of field
theories became an issue with the advent of early theories of the grand
unification of strong, electromagnetic and weak interactions, these GUT's being
renormalizable but decidedly unnatural. For a brief introduction to such
theories see Quigg's book:
 \begin{Ref} C.\ Quigg, {\em Gauge Theories of Strong, Weak, and Electromagnetic
Interactions}, Benjamin/Cummings, Reading, 1983.
 \end{Ref}

Nonrelativistic QED is practically as old as quantum mechanics, but the idea of
using this theory as a cut-off field theory, thereby permitting
rigorous calculations beyond tree level, originates with the (too) short paper
 \begin{Ref} W.E.\ Caswell and G.P.\ Lepage, {\em Phys.\ Lett.\/} {\bf 167B},
437(1986).
 \end{Ref}
Here the theory is applied in state-of-the-art bound-state calculations for
positronium and muonium. The application of these ideas to heavy-quark physics
is discussed in
 \begin{Ref} G.P.\ Lepage and B.A.\ Thacker, in {\em Field Theory on the
Lattice}, edited by A.\ Billoire {\em et al.\/}, {\em Nucl.\ Phys.\ (Proc.\
Suppl.)\/} {\bf 4} (1988);
 B.A.\ Thacker, Ph.D. Thesis, Cornell University (September 1989);
 B.A.\ Thacker and G.P.\ Lepage, Cornell preprint (December, 1989).
 \end{Ref}

 \enddoc